\documentclass{article}
\usepackage{spconf,amsmath,graphicx}
\usepackage{caption}
\usepackage{subcaption}
\usepackage{multirow}
\usepackage{color,soul}
\usepackage{placeins}

\title{ A lightweight convolutional neural network for image denoising with fine details preservation capability }
\name{Sutanu Bera, Avisek Lahiri, Prabir Kumar Biswas}
\address{Department of Electronics and Electrical Communication Engineering, \\Indian Institute of Technology Kharagpur
}
\begin{document}
%
\maketitle
\begin{abstract}
Image denoising is a fundamental problem in image processing whose primary objective is to remove the noise while preserving the original image structure. In this work, we proposed a new architecture for image denoising. We have used several dense blocks to design our network. Additionally, we have forwarded feature extracted in the first layer to the input of every transition layer. Our experimental result suggests that the use of low-level feature helps in reconstructing better texture. Furthermore, we had trained our network with a combination of MSE and a differentiable multi-scale structural similarity index(MS-SSIM). With proper training, our proposed model with a much lower parameter can outperform other models which were with trained much higher parameters. We evaluated our algorithm on two grayscale benchmark dataset BSD68 and SET12. Our model had achieved similar PSNR with the current state of the art methods and most of the time better SSIM than other algorithms.

\end{abstract}
\begin{keywords}
Image Denoising, MS-SSIM, Skip Connections, Low-level feature, Texture Preservation, CNN
\end{keywords}
\section{Introduction}
\label{sec:intro}

Image denoising is a fundamental image reconstruction problem, but still an active topic for low-level vision researchers. The main objective of image denoising is to recover the clean latent image $f$ from the noise corrupted version $g$, which follows the image degradation model $g= f+ \eta$, where $\eta$ is the additive noise. It has been shown that a efficient denoising algorithm can solve many other image reconstruction problems such as super resolution, deblurring, inpainting, compression etc.\cite{DBLP:journals/corr/RomanoEM16} 

Initially, learning the image prior was considered as the effective way of denoising the image. In particular, models like non-local similarity (NSS) models\cite{buades2005non}, Markov random field (MRF) models\cite{li2009markov}, Sparse models\cite{elad2006image}, etc. was used for denoising. \\
However, all the previous models require complex iterative computation, thus becomes time-consuming in the testing stage. On the contrary, for a real-time application, fast algorithms are needed. As a solution , discriminative models\cite{schmidt2014shrinkage} attracted attention because these models eliminated the iterative steps required at the testing stage.
\\In recent times, as a discriminative model, deep convolutional network\cite{mao2016image},\cite{zhang2018ffdnet},\cite{remez2017deep},\cite{8451132} has started gaining considerable attention because of its fast and efficient denoising capability. In this regard Kai Zhang et al.\cite{zhang2017beyond} proposed DnCNN network, they first utilized batch normalization and residual learning for denoising task. 
\begin{figure}
       \begin{subfigure}[b]{0.25\textwidth}
                \includegraphics[width=\linewidth]{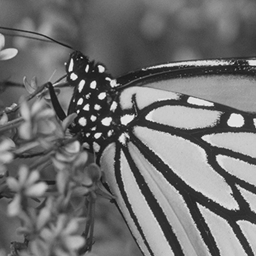}
                \caption{Original}
            \label{fig:or7}
        \end{subfigure}
       \begin{subfigure}[b]{0.25\textwidth}
                \includegraphics[width=\linewidth]{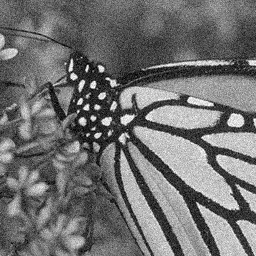}
                \caption{Noisy}
                \label{fig:n07}
        \end{subfigure}%
        
        \begin{subfigure}[b]{0.25\textwidth}
                \includegraphics[width=\linewidth]{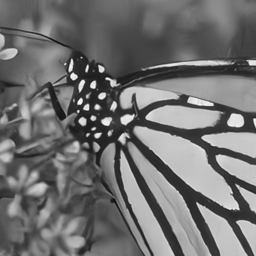}
                \caption{DnCNN(0.55M parameters)}
                \label{fig:m07}
        \end{subfigure}
        \begin{subfigure}[b]{0.25\textwidth}
                \includegraphics[width=\linewidth]{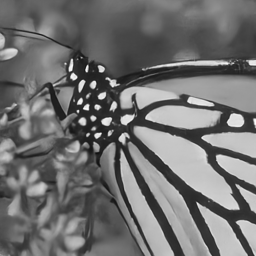}
                \caption{Proposed(0.13M parameters)}
                \label{fig:d07}
        \end{subfigure}
        \caption{ \footnotesize	Denoising Result on noise level $
        \sigma$ = 25, (a) Original, (b) Noisy (PSNR:24dB/SSIM:0.60) (c)DnCNN (PSNR:30.38dB/SSIM:0.91) (d) Proposed (PSNR:30.29dB/SSIM:0.91). Proposed model with $76\%$ lesser parameter than current state of art\cite{zhang2017beyond} yielded visually similar image. }
        \label{fig:parrot}
\end{figure}
Their model otperformed many of benchmark model such as; BM3D\cite{dabov2006image}, TNRD\cite{chen2017trainable}, etc. 

Indeed, Deep CNN based models provide better performance than other models. However, these models comprise of a good deal of matrix multiplication. Therefore, good computing resources are required for implementation. Also, attempt to reduce the complexity of network results in a reduction in performance. In this study, we tried to make a tradeoff between the model complexity and performance.\\
Meanwhile, Mean Square Error (MSE) or L2 norm of the difference image remains the most preferred error measure in the researcher community; it may be because L2 norm is convex, differentiable, easier to optimize. Additionally, it also provides the maximum likelihood (ML) estimate solution. Still, as an error measure, the L2 norm ignores the local characteristic of the image like, contrast, luminance, structural information, etc., which are an integral part of the human visual system. In this study, we also tried to incorporate this measures into loss function to train our network.
\\ \textbf{Contribution: }The main contributions of this study are as follows:
\begin{itemize}
    \item We present a new CNN architecture for image denoising with lesser complexity than the benchmark architectures. We also propose an optional addition to the above mentioned architecture to further reduce the complexity and execution time. Our model with lesser number of parameter produced similar result like more complex deep CNN models. 
    \item We forwarded the feature extracted in the first layer to next layer through skip connections. Our experimental result shows that, these low-level features contain complementary information that high level feature lacks for texture generation. Our solution reconstructs better texture than baseline solution.
    \item We also showed the effect of inclusion of local characteristic into the cost function by using MS-SSIM. We also propose specific sequence of multi stage training to generate optimal solution.
\end{itemize}
Remaining part of the paper is organized in the following way. In Section 2 the proposed architecture is introduced, then in Section 3 the training procedure and the cost function is explained, and in Section 4 the evaluation results are represented, and finally, Section 5 provides the conclusion about the findings of this study. 
\begin{figure}
    \centering
    \includegraphics[width= 0.5\textwidth, height=1in]{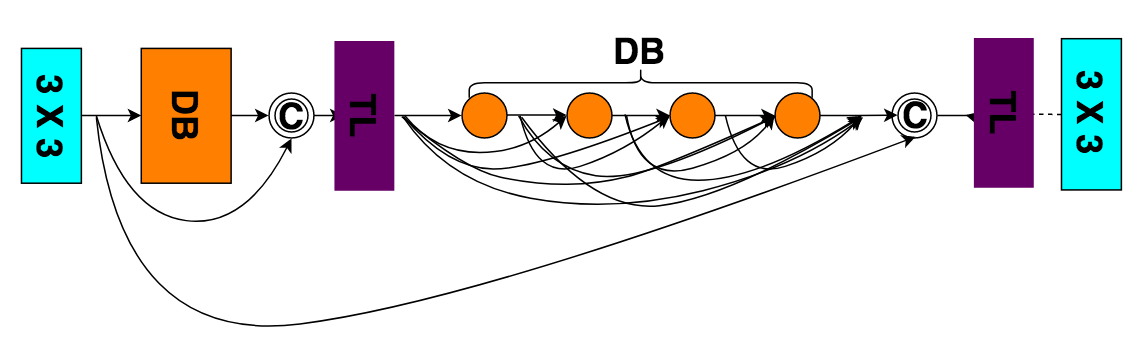}
    \caption{Proposed Architecture DB: Dense Block, TL: Transition Layer}
    \label{fig:model}
\end{figure}
\section{Proposed Architecture}
It is known that the initial layer of deep neural network learns low-level feature like edge information, corner points, etc, whereas the deeper layer learns more complex feature like face orientation, larger shapes, etc. Perhaps, these high-level features are useful in classification task, but for image restoration these features might become less significant. Indeed, the low-level features plays an important role in image restoration. With this foundation, we decided to propagate low-level features through the network and let the network infer the importance of these features for reconstruction. Figure \ref{fig:model} illustrates the proposed network. First layer is a convolution layer with 64 filters of $3\times3$ kernel size. Next, a series of dense block (DB)and transition layer (TL) is used to extract features, and then finally a convolution layer with $3\times3$ kernel to reconstruct the image. We used total 6 pairs of dense block and transition layer in our network.
Inside the DB, the feature-maps of all preceding convolution layers are used as the input to current layer, and its own feature-maps are used as inputs into all subsequent layers. The growth rate of every DB was set to 16, and four convolution layer followed by batch normalization and ReLU non-linearity was used in each DB. After every dense block, a transition layer consisting $1\times1$ convolution layer followed by batch normalization and ReLU, had been used to reduce the depth of the feature map and, also to combine the feature extracted at different layer. According to the NIN paper\cite{lin2013network}, $1\times1$ convolution is similar to cross-channel parametric pooling. This cascaded cross channel parametric pooling structure allows complex and learnable interactions of cross channel information. The input of the transition layer is a mixture of the low-level and high-level feature over the volume. The $1\times1$ convolution may drop the low-level feature or can propagate them to the next layer.   
\\Adapting dense net\cite{huang2017densely} architecture helped in reusing the feature map. As a result, the total number of parameter reduced significantly. Furthermore, a better receptive field was obtained by using more number of layers with fewer filters. Additionally, the use of skip connection also makes the error surface more smooth and convex\cite{li2018visualizing}. In addition to this, traditional convolution can also be replaced with depthwise separable convolution\cite{howard2017mobilenets} to further reduce the number of parameters drastically. 
\begin{table*}[]
\centering
\resizebox{\textwidth}{!}{%
\begin{tabular}{|c|c|c|c|c|c|c|c|c|c|c|}
\hline
Dataset & Noise Level & \multicolumn{9}{c|}{Method} \\ \hline
 &  & BM3D\cite{dabov2006image}  & WNNM\cite{gu2014weighted} & EPLL\cite{zoran2011learning} & MLP\cite{burger2012image} & CSF\cite{schmidt2014shrinkage} & TNRD\cite{chen2017trainable} & DnCNN\cite{zhang2017beyond}  & FFDNet\cite{zhang2018ffdnet} & Proposed v1 \\ \hline
 \hline
\multirow{3}{*}{BSD68} & 15 & 31.07 & 31.37 & - & 31.24 & 31.42 &31.73 &  \textcolor{red}{31.75} & 31.63 & \textcolor{green}{31.70} \\ \cline{2-11} 
 & 25 & 28.57 & 28.83 & 28.68 & 28.96 & 28.74 & 28.92 & \textcolor{red}{29.23} & 29.19 & \textcolor{green}{29.20} \\ \cline{2-11} 
 & 50 & 25.62 & 25.87 & 25.67 & 26.03 & - & 25.97 & 26.23 & \textcolor{red}{26.29} & \textcolor{green}{26.25} \\ \hline
 \hline
\multirow{3}{*}{SET12} & 15 & 32.37 & 32.69 & 32.13 & - & 32.31 & 32.50 & \textcolor{red}{32.88} & 32.75 & \textcolor{green}{32.82} \\ \cline{2-11} 
 & 25 & 29.96 & 30.25 & 29.69 & 30.02 & 29.83 & 30.05 & \textcolor{red}{30.45} & \textcolor{green}{30.43} & 30.41 \\ \cline{2-11} 
 & 50 & 26.72 & 27.05 & 26.47 & 26.78 & - & 26.81 & \textcolor{green}{27.23} & \textcolor{red}{27.32} & 27.17 \\ \hline

\end{tabular}%
}
\caption{The average PSNR(dB) results of different methods, top 2 results are marked in red and green colour respectively.}
\label{psnr}
\end{table*}
\begin{figure*}
        \begin{subfigure}[b]{0.25\textwidth}
                \includegraphics[width=\linewidth]{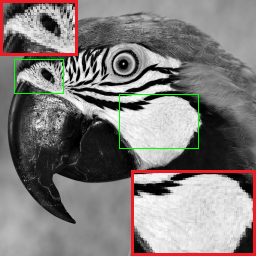}
                \caption{Original}
                \label{fig:or7}
        \end{subfigure}%
        \begin{subfigure}[b]{0.25\textwidth}
                \includegraphics[width=\linewidth]{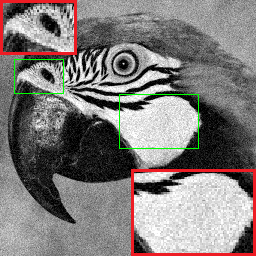}
                \caption{Noisy}
                \label{fig:n07}
        \end{subfigure}%
        \begin{subfigure}[b]{0.25\textwidth}
                \includegraphics[width=\linewidth]{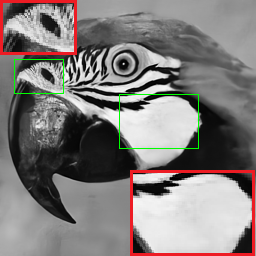}
                \caption{Using Low-level feature}
                \label{fig:m07}
        \end{subfigure}%
        \begin{subfigure}[b]{0.25\textwidth}
                \includegraphics[width=\linewidth]{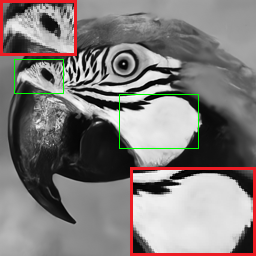}
                \caption{Without low-level feature}
                \label{fig:d07}
        \end{subfigure}
        \caption{\footnotesize	Denoising Result of Parrot image on noise level $
        \sigma$ = 25. (The reader is encouraged to zoom in for a better view). Reconstructed image using low-level feature has better texture than the other solution. (a) Original, (b) Noisy, (c) Proposed v1 (d) We removed the skip connection before all the transition layer so that low-level feature does not propagate through the network. } \label{fig:parrot}
\end{figure*}

\begin{figure*}[htb]
    \centering 
\begin{subfigure}{0.32\textwidth}
  \includegraphics[width=\linewidth]{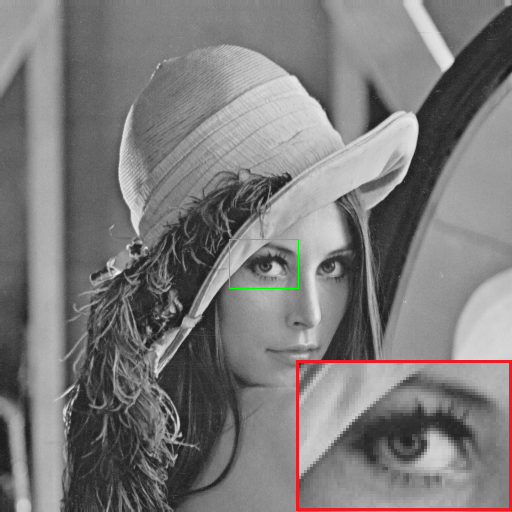}
  \caption{Original}
  \label{fig:1}
\end{subfigure}\hfill
\begin{subfigure}{0.32\textwidth}
  \includegraphics[width=\linewidth]{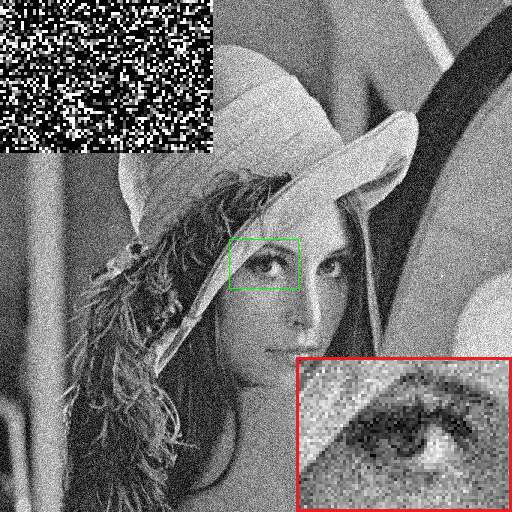}
  \caption{Noisy}
  \label{fig:2}
\end{subfigure}\hfill 
\begin{subfigure}{0.32\textwidth}
  \includegraphics[width=\linewidth]{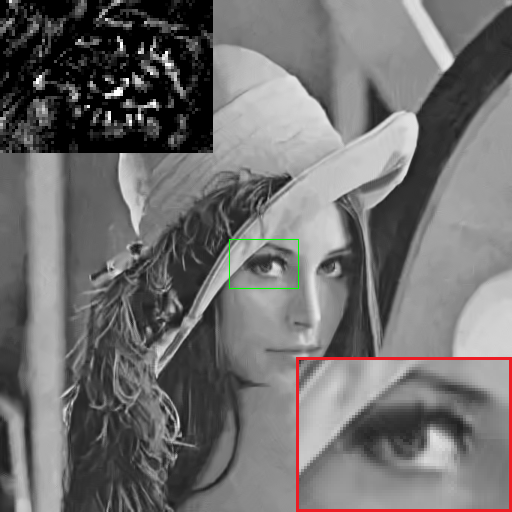}
  \caption{BM3D}
  \label{fig:3}
\end{subfigure}
\medskip
\begin{subfigure}{0.32\textwidth}
  \includegraphics[width=\linewidth]{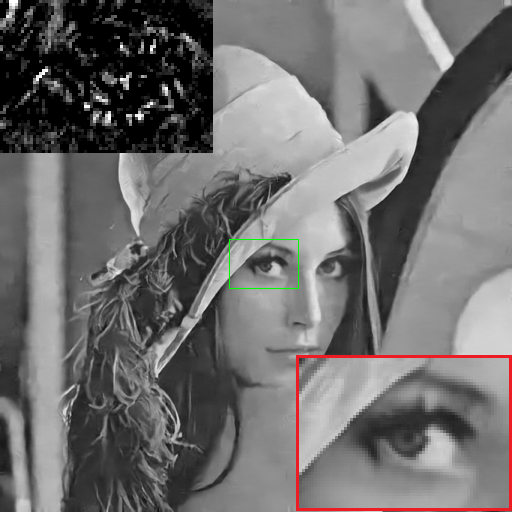}
  \caption{TNRD}
  \label{fig:4}
\end{subfigure}\hfill 
\begin{subfigure}{0.32\textwidth}
  \includegraphics[width=\linewidth]{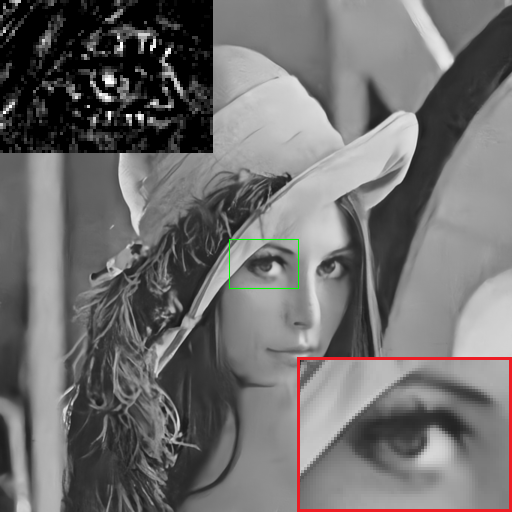}
  \caption{DnCNN}
  \label{fig:5}
\end{subfigure}\hfill
\begin{subfigure}{0.32\textwidth}
  \includegraphics[width=\linewidth]{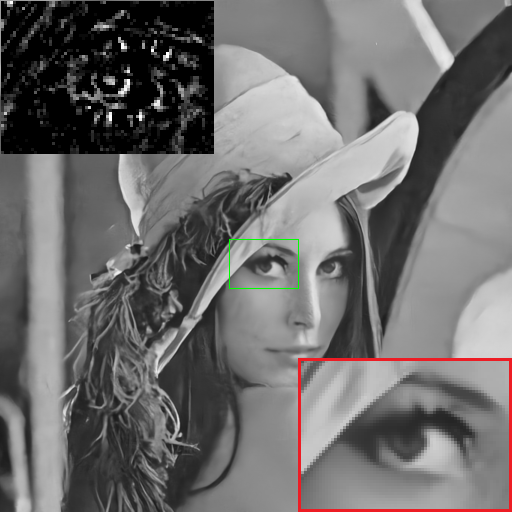}
  \caption{Proposed v1}
  \label{fig:6}
\end{subfigure}
\caption{\footnotesize	Denoising result on lena image with noise level 25 by different method. Proposed method performed better in preserving fine details while removing noise. The lower inset is the zoomed version of marked area, and the upper inset is the difference image of same region.}
\label{fig:method}
\end{figure*}

\section{Training Details}
Use of residual learning in image reconstruction is already established\cite{zhang2017beyond}. According \cite{he2016deep} to when the original mapping is more similar to identity mapping then it is more easy to learn and optimize the residual mapping. Then clean image can restored by subtracting the residual map(i.e. the noise map) from the noisy image.
\\Now, one obvious choice for error measure is L2 norm of difference image or the Mean Square Error. However, L2 does not consider the local characteristic of the image. To overcame this drawback, we first trained the entire network using MSE. Then, retrained only the last layer of the network with the combination L2 norm and differentiable Multi Scale Structural Similarity Index (MS-SSIM) as a multi objective optimization problem keeping other layer untouched. Why we only trained the last layer only with MS-SSIM is discussed in the Section 4. \\The cost function for training the final layer of the image is given by 
{\[\ l(\Theta) =(1- MS-SSIM(x,(y-R(y;\Theta))) \]}
{\[\ + \frac{1}{2N} \sum_{i=1}^{N} || R(y_{i}; \Theta) - (y_{i} - x_{i})||_{F}^{2} \]}
The differentiable implementation of MS-SSIM can be found in the study by Hang Zhao \textit{et al.}\cite{zhao2017loss}. 
\\We trained our network for three specific noise level, particularly for $\sigma =$ 15, 25, 50. For validation two benchmark dataset, Berkeley segmentation data set (BSD68) containing 68 natural images, and famous 12 images of SET12, is used. 

\section{Result and Discussion}


We evaluated both of our model namely; proposed v1: as described above section, and proposed v2: replacing the convolution layer by depthwise separable convolution. Table \ref{psnr} depicts the comparison of the proposed v1 network with other method applied in this two dataset. It can be seen that proposed method has outperformed many benchmark algorithm like BM3D, TNRD, WNMM, CSF and also yielded comparable PNSR with DnCNN algorithm.
\begin{table}[]
\centering
\resizebox{0.45\textwidth}{!}{%
\begin{tabular}{cccccc}
 &  & \multicolumn{4}{c}{METHOD} \\ \hline
\multicolumn{1}{|c|}{Dataset} & \multicolumn{1}{c|}{Noise Level} & \multicolumn{1}{c|}{BM3D} & \multicolumn{1}{c|}{TNRD} & \multicolumn{1}{c|}{DnCNN} & \multicolumn{1}{c|}{Proposed v1} \\ \hline
\multicolumn{1}{|c|}{\multirow{3}{*}{BSD 68}} & \multicolumn{1}{c|}{$\sigma$=15} & \multicolumn{1}{c|}{0.8741} & \multicolumn{1}{c|}{0.8947} & \multicolumn{1}{c|}{0.8947} & \multicolumn{1}{c|}{\textbf{0.8949}} \\ \cline{2-6} 
\multicolumn{1}{|c|}{} & \multicolumn{1}{c|}{$\sigma$=25} & \multicolumn{1}{c|}{0.8025} & \multicolumn{1}{c|}{0.8206} & \multicolumn{1}{c|}{0.8321} & \multicolumn{1}{c|}{\textbf{0.8326}} \\ \cline{2-6} 
\multicolumn{1}{|c|}{} & \multicolumn{1}{c|}{$\sigma$=50} & \multicolumn{1}{c|}{0.6744} & \multicolumn{1}{c|}{0.7104} & \multicolumn{1}{c|}{0.7252} & \multicolumn{1}{c|}{\textbf{0.7289}} \\ \hline
\multicolumn{1}{|c|}{\multirow{3}{*}{SET 12}} & \multicolumn{1}{c|}{$\sigma$=15} & \multicolumn{1}{c|}{0.8989} & \multicolumn{1}{c|}{0.9004} & \multicolumn{1}{c|}{0.9073} & \multicolumn{1}{c|}{\textbf{0.9099}} \\ \cline{2-6} 
\multicolumn{1}{|c|}{} & \multicolumn{1}{c|}{$\sigma$=25} & \multicolumn{1}{c|}{0.8553} & \multicolumn{1}{c|}{0.8573} & \multicolumn{1}{c|}{0.8675} & \multicolumn{1}{c|}{\textbf{0.8678}} \\ \cline{2-6} 
\multicolumn{1}{|c|}{} & \multicolumn{1}{c|}{$\sigma$=50} & \multicolumn{1}{c|}{0.7679} & \multicolumn{1}{c|}{0.7753} & \multicolumn{1}{c|}{\textbf{0.7913}} & \multicolumn{1}{c|}{\textbf{0.7913}} \\ \hline
\end{tabular}%
}
\caption{Comparision of SSIM with other state of art models.}
\label{ssim}
\end{table}
\\Figure \ref{fig:parrot} gives an example to show the effect of adding low-level feature to the input of every transition layer. Reconstructing the texture or the fine details is one of the most challenging tasks for image denoising models. Indeed, most of the algorithm produces a smoothed version of the image. However, it can be seen that adding low-level feature provides better texture. In Figure \ref{fig:parrot} the fine texture near the nose and also in the area near chin is noticable. Table \ref{ssim} reports comparative SSIM of our model with other three benchmark algorithm.
\begin{table}[]
\centering
\resizebox{0.4\textwidth}{!}{%
\begin{tabular}{|c|c|c|c|c|}
\hline
\textbf{Dataset} & \textbf{Noise Level} & \textbf{DnCNN} & \textbf{Proposed v1} & \textbf{Proposed v2}\\ \hline
\hline
\multicolumn{1}{|l|}{\multirow{3}{*}{BSD68}} & 15 & 31.75 & 31.70 & 31.62 \\
\multicolumn{1}{|l|}{} & 25 & 29.23 & 29.20 & 29.08 \\
\multicolumn{1}{|l|}{} & 50 & 26.23 & 26.25 & 26.12 \\ \hline
\hline
\multirow{3}{*}{SET12} & 15 & 32.88 & 32.82 & 32.69 \\ 
 & 25 & 30.45 & 30.41 & 30.32 \\
 & 50 & 27.23 & 27.17 & 27.09 \\ \hline
 \hline
Parameter &  & 556032 & 382080 & 133248 \\ \hline
\end{tabular}%
}
\caption{Comparison of Parameter and PSNR(dB)}
\label{para}
\end{table}
In most of the times, We had scored best SSIM. Retraining last layer with MS-SSIM helped in achieving this. Furthermore, we also witnessed a small improvement in PSNR after training with MS-SSIM.  We also tried to retrain all the layer with MS-SSIM and, but doing so resulted in similar SSIM but lower PSNR images.  
\\ Table \ref{para} shows comparison between the number of trainable parameter and PSNR. Model trained with depthwise separable convolution has nearly $76\%$ less parameter than the DnCNN model, but has achieved similar PSNR, also our model with normal convolution has $30\%$ less parameter than DnCNN.

The visual comparisons of different methods are given in Figure \ref{fig:method}. Proposed model has kept fine details better than the other model. The shape of the reconstructed eye is better our model, it can be also verified by examining the difference image. Overall, proposed solution provides good perceptual quality image.

\section{Conclusion}
 In this paper a new CNN network for image denoising is proposed. This method has less complexity than the state of the art method, but still resultant images are similar with better texture. This light weight model can be utilised in low resources devices, such as smart phones, to perform state of art denoising.
\bibliographystyle{IEEEbib}
\bibliography{strings,refs}

\end{document}